\documentclass[preprint,12pt]{article}
\usepackage{amssymb}
\usepackage{amsmath}
\usepackage{hyperref}

\begin{document}

\centerline{}

\centerline {\Large{\bf On the nonlocal Darboux transformation  
 }}

\centerline{}

\centerline{\Large{\bf for the stationary axially symmetric   }}

\centerline{}

\centerline{\Large{\bf Schr\"odinger and Helmholtz equations}}

\centerline{}

\centerline{\bf {A. G. Kudryavtsev}}

\centerline{}

\centerline{Institute of Applied Mechanics,}

\centerline{Russian Academy of Sciences, Moscow 125040, Russia}

\centerline{kudryavtsev\_a\_g@mail.ru}

\begin{abstract}

The nonlocal Darboux transformation for the stationary axially symmetric 
Schr\"odinger and Helmholtz equations is considered. Formulae for the 
nonlocal Darboux transformation are obtained and its relation to the 
generalized Moutard transformation is established.
New examples of two - dimensional  potencials and exact solutions
for the stationary axially symmetric Schr\"odinger and 
Helmholtz equations are obtained as an application 
of the nonlocal Darboux transformation.

\end{abstract}

\centerline{PACS: 02.30.Jr, 02.30.Ik, 03.65.Ge}






\section{Introduction}

Consider the stationary Schr\"odinger equation in the form
\begin{equation*}
\left( \Delta -{\it u}\left( x,y,z \right) \right) Y \left( x,y,z
\right) =0 \,.
\end{equation*}
In the case ${\it u}=-E+V\left( x,y,z \right)$ this equation 
describes nonrelativistic quantum system with energy $E$ \cite
{Landau}. In the case ${\it u}= -{\omega} ^{2}/{c\left( x,y,z
\right)}^{2}$ equation describes an acoustic pressure
field with temporal frequency $\omega$ in inhomogeneous media with
sound velocity $c$ \cite {Morse} and is known as Helmholtz equation
 \cite {Vinogradova}.  The case of fixed frequency
$\omega$ is of interest for modelling in acoustic tomography \cite
{Kak}. The case of fixed energy $E$ for two - dimensional equation 
is of interest for the multidimentional inverse scattering theory 
due to connections with two - dimensional integrable nonlinear 
systems \cite {Novikov2010}.

In the case of axial symmetry  the stationary equation 
 in cylindrical coordinates has the form

\begin{equation} \label{eq1}
 \left({\frac {\partial ^{2}}{\partial {r}^{2}}} +{\frac {1}{r}} 
{\frac {\partial }{\partial r}}+{\frac {
\partial ^{2}}{\partial {z}^{2}}}-u \left( r,z
 \right)  \right)Y \left( r,z \right) = 0 \,.
\end{equation}
The Darboux-type transformations are useful for obtaining 
new solvable linear differential equations from initial solvable equation.
The authors of the book \cite {Matveev1991} note that the importance
of the Darboux transformation lies in the possibility of obtaining new solvable
equations based on the original solvable equation. Changing the solution of
the original equation in the Darboux transformation formula gives solutions
to the new equation. If a set of selected special functions allows representing
all solutions analytically, we can say that the original equation is exactly solvable.
If the original equation is exactly solvable, you can get all the solutions of the
new equation using the Darboux transform. The Darboux transformation can be
repeated any number of times. Thus, we can say that an equation is solvable if
it can be obtained from another solvable equation by a finite number of Darboux
transformations. For example, if the initial equation is the stationary Schr\"odinger
equation in one-dimensional quantum mechanics, it is natural to call the equations
describing the free motion of particles, the harmonic oscillator, and the Coulomb and
Morse potentials solvable. In this paper, we say that the stationary Schr\"odinger
and Helmholtz equations are solvable in the two-dimensional case if they can be
obtained from an equation with a constant (including zero) potential by using
a finite number of Darboux transformations.

Applications of the classical Darboux transformation for the 
one - dimensional Schr\"odinger equation can be found in 
the book  \cite {Matveev1991}. The classical Moutard 
transformation for the two - dimensional Schr\"odinger equation 
in cartesian coordinates is reviewed in  \cite {Athorne 1991}.
Various generalizations of the classical Darboux transformations  
and their applications to the  two - dimensional systems
were investigated, see  review \cite {Andrianov 2012}, recent publication
\cite {Ioffe  2020} and references therein.
In the papers \cite {Kudryavtsev2013},  \cite {Kudryavtsev2016}
the nonlocal variable was included in Darboux transformation. 
The nonlocal Darboux transformation of the two - dimensional
stationary Schr\"odinger equation in cartesian coordinates was 
obtained and its relation to the Moutard transformation was 
established. The main idear of this papers was inspired by the 
approach to nonlocal symmetries in the symmetry group analysis 
of differential equations  \cite {Ibragimov1991},  
\cite {Bluman2010} (for the variant based on the theory of coverings, 
see \cite { Vinogradov 1999} and references therein) .
In the paper \cite {Kudryavtsev2020} the generalized Moutard transformation 
was considered and applied for the Schr\"odinger equation in cylindrical coordinates.
In the present paper we consider the nonlocal Darboux transformation 
for the stationary Schr\"odinger equation in cylindrical coordinates \eqref{eq1}
using the approach of papers \cite {Kudryavtsev2013},  
\cite {Kudryavtsev2016}.

We use the relation of the Schr\"odinger equation with the 
Fokker-Planck equation  \cite {Risken1989}. By the substitution
\begin{equation} \label{eq2}
Y \left( r,z \right) =P \left( r,z \right) {e^{h \left( r,z
\right) }}
\end{equation}
we obtain the Fokker-Planck type equation
\begin{equation} \label{eq3}
{\frac {\partial }{\partial r}} \left( {\it P_{r}}
+2\,{ \it h_{r}}\,P+{\frac {1}{r}}\,P \right)+{\frac {\partial }{\partial z}} \left(
{\it P_{z}} +2\,{\it h_{z}}\,P \right) =0
\end{equation}
if $u$ and $h$ satisfy the condition
\begin{equation} \label{eq4}
{\it u}=-{\it h_{rr}} +{{\it h_{r}}}^{2}+{\frac {1}{r}}\, {\it h_{r}}+{\frac {1}{{r}^{2}}} 
-{\it h_{zz}} +{{\it h_{z}}}^{2} \,.
\end{equation}

The equation \eqref{eq3} has the conservation law
form that yields a pair of equations

\begin{equation} \label{eq5}
{\it P_{r}}+ 2\,{ \it h_{r}}\,P+{\frac {1}{r}}\,P  -{\it Q_{z}} =0 \,,
\end{equation}
\begin{equation} \label{eq6}
{\it P_{z}}+ 2\,{ \it h_{z}}\,P  +{\it Q_{r}} =0 \,.
\end{equation}

The variable $Q$ is a nonlocal variable for the equation \eqref{eq3}
and related Schr\"odinger equation \eqref{eq1}.
The Darboux transformation for the equations \eqref{eq5},
\eqref{eq6} including  $Q$ provides the nonlocal Darboux transformation for the 
Schr\"odinger equation  \eqref{eq1}. 

\section{The nonlocal Darboux transformation }

Let us consider linear operator corresponding to the system of
equations \eqref{eq5}, \eqref{eq6}

\begin{equation*}
\hat L\left( h \left( r ,z \right)  \right) \, {\bf f}=
\begin{pmatrix} { 2\,{ \it h_{r}}+{\frac {1}{r}}+\frac {\partial }{\partial r} }  &
{ -\frac {\partial }{\partial z} } \\ {{2\,{ \it h_{z}}+\frac
{\partial }{\partial z}}}  & {\frac {\partial }{\partial r}}
\end{pmatrix}\,
\begin{pmatrix} f_1 \\f_2 \end{pmatrix} \,.
\end{equation*}

Consider Darboux transformation in the form

\begin{equation*}
\hat L_D \, {\bf f}=
\begin{pmatrix} { g_{11}-a_{11}\,\frac {\partial }{\partial r}-b_{11}\,\frac {\partial }{\partial z}  }  &
{  g_{12}-a_{12}\,\frac {\partial }{\partial r}-b_{12}\,\frac
{\partial }{\partial z} } \\ { g_{21}-a_{21}\,\frac {\partial
}{\partial r}-b_{21}\,\frac {\partial }{\partial z} }  & {
g_{22}-a_{22}\,\frac {\partial }{\partial r}-b_{22}\,\frac
{\partial }{\partial z} }
\end{pmatrix} \,
\begin{pmatrix} f_1 \\ f_2 \end{pmatrix} \,.
\end{equation*}

If linear operators $\hat L$ and $\hat L_{D}$ hold the
intertwining relation

\begin{equation} \label{eq7}
\left( \hat L\left( h \left( r ,z \right) + s \left( r ,z
\right) \right)\hat L_{D} - \hat L_{D} \hat L\left( h \left( r ,z
\right) \right) \right) \, {\bf f}= 0
\end{equation}
for any $ {\bf f} \in  \mathcal{F} \supset Ker\left( \hat L\left(
h \right)\right)$ where  $Ker\left( \hat L\left( h
\right)\right)=\{{\bf f}:{\hat {L}}\left( h \right){\bf f}=0\}$,
then for any ${\bf f_s}\in Ker\left( \hat L\left( h
\right)\right)$
 the function $\tilde {\bf f} \left( r,z \right)=
\hat L_{D} {\bf f_s} \left( r,z \right)$ is a solution of the
equation ${\hat {L}}\left( {\tilde {h}} \right) \tilde {\bf f} =0$
\, with the new potential $\tilde h = h+s$.

The equations for $s, g_{ij}, a_{ij}, b_{ij}$ can be obtained
from the intertwining relation  \eqref{eq7}. 

When solving equation \eqref{eq7} the following 
expression arises:
\begin{equation} \label{eq8}
V \left( r,z \right) =s \left( r,z \right) +2\,h \left( r,z \right) +
\ln  \left( r \right) \,.
\end{equation}
The special situation occur 
in the case $V \left( r,z \right) = 0$. In this case we obtain 
from the equation \eqref{eq7} the following transformation

\begin{equation} \label{eq9}
\hat L_D = r\,{e^{2\,h \left( r,z \right) }}
\begin{pmatrix} { 0 }  &
{ 1 } \\ { - 1 }  & { 0 }
\end{pmatrix}  \,.
\end{equation}

This transformation corresponds to the generalisation of
Moutard transformation.  The generalized Moutard transformation 
was applied for the Schr\"odinger equation in cylindrical coordinates 
in the paper \cite {Kudryavtsev2020}. Here we write out formulae
of generalized Moutard transformation for the sake of completeness
and subsequent use.

By the formula \eqref{eq4} 
with $\tilde h = -h  -\ln  \left( r \right)$ we 
obtain for the new Schr\"odinger potential

\begin{equation} \label{eq10}
\tilde {u} \left( r ,z \right) = u\left( r ,z \right) 
+2\,{\frac {\partial ^{2}}{\partial {r}^{2}}}h \left( r,z \right) 
+2\,{\frac {\partial ^{2}}{\partial {z}^{2}}}h \left( r,z \right) 
-{\frac {1}{{r}^{2}}}  \,.
\end{equation}

Consider
\begin{equation} \label{eq11}
Y_h\left( r ,z \right)={\frac {1}{r}}\,{{\rm e}^{-h \left( r,z \right) }}  \,.
\end{equation}

According to the formula \eqref{eq4}, $Y_h$ is a solution of the
Schr\"odinger equation  \eqref{eq1} with potential $u$.
Then we get another form of the formula \eqref{eq10}

\begin{equation} \label{eq12}
\tilde {u} \left( r ,z \right) = u\left( r ,z \right) 
-2\,{\frac {\partial ^{2}}{\partial {r}^{2}}}\ln  \left( {\it Y_h}
 \left( r,z \right)  \right) +{\frac {1}{{r}^{2}}} 
-2\,{\frac {\partial ^{2}}{\partial {z}^{2}}}\ln  
\left( {\it Y_h} \left( r,z \right)  \right)  \,.
\end{equation}

This formula is the generalization of the formula 
of Moutard transformation for the potential of the 
Schr\"odinger equation.

From the formula \eqref{eq9} and the relation

\begin{equation} \label{eq13}
\tilde {Y} \left( r,z \right) =\tilde {P} \left( r,z \right) {e^{
\tilde {h} \left( r,z \right) }}
\end{equation}

we have the following equations

\begin{equation} \label{eq14}
\tilde {P} \left( r ,z \right) = r\,{e^{2\,h \left( r,z \right) }}Q
\left( r,z \right)  \,,
\end{equation}

\begin{equation} \label{eq15}
\tilde {Y} \left( r,z \right) = {e^{h \left( r,z \right) }}Q
\left( r,z \right)  \,.
\end{equation}

One can express $P, Q, h$ by equations \eqref{eq2}, \eqref{eq15},
\eqref{eq11} trough $Y, \tilde {Y}, Y_h$ and substitute to the
system of equations \eqref{eq5}, \eqref{eq6}. The result is

\begin{equation} \label{eq16}
{\frac {\partial }{\partial z}} \left( {\it Y_h} \left( r,z
\right) {\tilde {Y}} \left( r,z \right)  \right) - \left( {\it
Y_h} \left( r,z \right)  \right) ^{2}{\frac {\partial }{\partial
r}} \left( {\frac {{ \it Y} \left( r,z \right) }{{\it Y_h} \left(
r,z \right) }} \right)=0  \,,
\end{equation}

\begin{multline} \label{eq17}
{\frac {\partial }{\partial r}} \left( {\it Y_h} \left( r,z
\right) {\tilde {Y}} \left( r,z \right)  \right) 
+{\frac {1}{r}}\,{\it Y_h} \left( r,z \right) {\tilde {Y}} \left( r,z \right)
\\
+ \left( {\it Y_h} \left( r,z \right)  \right) ^{2}{\frac {\partial }{\partial
z}} \left( {\frac {{ \it Y} \left( r,z \right) }{{\it Y_h} \left(
r,z \right) }} \right)=0  \,.
\end{multline}
Here ${ \it Y}, {\it Y_h}$ are the solutions of equation \eqref{eq1}
with the initial potential $u$, and the function $ {\tilde {Y}}$
defined as a solution of a consistent system of
equations \eqref{eq16} and \eqref{eq17} is a solution 
of equation \eqref{eq1} with the new potential $\tilde {u}$.

These formulae are the generalization of the formulae of Moutard
transformation for the solution of the Schr\"odinger equation.
Thus the case $V \left( r,z \right) = 0 \,$ of the 
nonlocal Darboux transformation for the stationary axially symmetric 
Schr\"odinger equation provides the generalization
of the Moutard transformation.

Note that 
${\it Y}={\it Y_h}$ provides ${\tilde {Y}}={\left(r\,{\it Y_h} \right)}^{-1} $
as the simple example of solution for equations 
\eqref{eq16}, \eqref{eq17}.

It is convenient to use the formula for the superposition of two
generalized Moutard transformations. Let ${\it Y_1}$ and ${\it Y_2}$ 
are solutions of equation \eqref{eq1} with potential $u$.
Applying consistently the formulas of the generalized
Moutard transformation, we obtain the following superposition formulas
for two transformations:

\begin{multline} \label{eq18}
{\tilde {\tilde {u}}}  =
u-2\,{\frac {\partial ^{2}\ln \left( F \right) }{\partial {r}^{2}}}
-2\,{\frac {\partial ^{2}\ln  \left( F \right) }{\partial {z}^{2}}}
\\
=u +2 \left( {\frac {\partial {\it Y_2} }{\partial z
}}  \right)  \left( 2\,r{\frac {\partial {\it Y_1}}{
\partial r}} +{\it Y_1} \right) F^{-1}
\\
-2 \left( {\frac {\partial {\it Y_1} }{\partial z
}}  \right)  \left( 2\,r{\frac {\partial {\it Y_2}}{
\partial r}} +{\it Y_2} 
 \right) F^{-1}  
\\
+2\,{r}^{2}   \left(  {\frac {\partial {\it Y_2} }{\partial r}}{\it Y_1}   -{
\it Y_2}  {\frac {\partial {\it Y_1} }{\partial r}}
 \right) ^{2} F^{-2}
\\
+2\,{r}^{2}   \left(  {\frac {\partial {\it Y_2} }{\partial z}}{\it Y_1}   -{
\it Y_2}  {\frac {\partial {\it Y_1} }{\partial z}}
 \right) ^{2} F^{-2}
\, ,
\end{multline}
where function $F$ satisfies the consistent system of equations
\begin{equation} \label{eq19}
{\frac {\partial }{\partial z}}F = r \left( 
 {\frac {\partial {\it Y_2}}{\partial r}}{\it Y_1} -{\it Y_2}{\frac {\partial {
\it Y_1}  }{\partial r}} \right)
\, ,
\end{equation}
\begin{equation} \label{eq20}
{\frac {\partial }{\partial r}}F =- r \left( 
 {\frac {\partial {\it Y_2}}{\partial z}}{\it Y_1} -{\it Y_2}{\frac {\partial {
\it Y_1}  }{\partial z}} \right)
\, .
\end{equation}
Formulas  \eqref{eq18} -  \eqref{eq20} are invariant under
substitution 
${\it Y_1} \rightarrow {\it Y_2}, {\it Y_2} \rightarrow {\it Y_1}, F  \rightarrow -F$. 
This reflects the commutativity property of generalized Moutard transformations,
the result does not depend on the order of choice of functions
${\it Y_1}, {\it Y_2}$.
Examples of solutions for the equation  \eqref{eq1} with potential  \eqref{eq18} 
can be obtained by the formulas
$ {\tilde {\tilde {Y}}}_{1} = {\it Y_1} F^{-1},\,  
{\tilde {\tilde {Y}}}_{2} = {\it Y_2}F^{-1}$.

In the general case, when $V$ is not equal to zero, we obtain from the equation
\eqref{eq7}  the following Darboux transformation

\begin{equation} \label{eq21}
\hat L_D = {e^{- s \left( r,z \right) }}
\begin{pmatrix} { R_{1}+\frac {\partial }{\partial z}  }  &
{  R_{2} } \\ { {s}_{r}-R_{2} }  & { {s}_{z}+R_{1}+ \frac
{\partial }{\partial z}}
\end{pmatrix} \,,
\end{equation}

where
\begin{gather*} 
{\it  R_{1}} ={\frac{1}{2}}\,\left({V}_{z} -2\,{s}_{z} + \left( {V}_{z} H+ {V}_{r} T\right) /{G }\right) \,,
\\
{\it  R_{2}} ={\frac{1}{2}}\,\left({s}_{r} +\left( {s}_{r} H- {s}_{z} T\right)/{G }\right) \,,
\\
{\it G}  ={s}_{r}{V}_{r}+{s}_{z}{V}_{z}, \,\,{\it H}  ={V}_{zz}-{s}_{rr}, \,\,{\it T}  ={V}_{rz}+{s}_{rz}
\end{gather*}
and $s$ satisfies the following system of two nonlinear differential
equations:
\begin{multline} \label{eq22}
G \left( G+H \right) s_{{{\it rz}}}-GTs_{{{\it zz}}} + \left( -G_{{z}}H
+ \left(H_{z}- V_{{r}}T \right) G \right) s_{{r}} \,
\\
\shoveleft{
+ \left( G_{{z}}T -  \left( V_{{z}}T+T_{{z}} \right) G \right) s_{{z}}
- \left( V_{{z}}H+V_{{r}}T \right) G_{{r}}+V_{{{\it rz}}}{G}^{2}
}
\\
+ \left( V_{{{\it rz}}}H
+V_{{z}}H_{{r}}+V_{{{\it rr}}}T+V_{{r}}T_{{r}} \right) G = 0 \,,
\end{multline}
\begin{multline} \label{eq23}
G \left( G+H \right) s_{{{\it rr}}}-GTs_{{{\it rz}}}+ \left( -G_{{r}}H
+V_{{r}}{G}^{2}+ \left( H_{{r}}+V_{{r}}H \right) G \right) s_{{r}}
\\
\shoveleft{
+
 \left( G_{{r}}T+{G}^{2}V_{{z}}+ \left( V_{{z}}H-T_{{r}} \right) G
 \right) s_{{z}}+ \left( V_{{z}}H+V_{{r}}T \right) G_{{z}}-V_{{{\it zz
}}}{G}^{2}
}
\\
- \left( V_{{{\it zz}}}H+V_{{z}}H_{{z}}+V_{{{\it rz}}}T+V_{
{r}}T_{{z}} \right) G = 0 \,.
\end{multline}
From the formula \eqref{eq21} and the relation \eqref{eq13} we have
\begin{equation} \label{eq24}
\tilde {P} = 
{{\rm e}^{-s}}
\left( {\it R_1} P  +{\frac {
\partial }{\partial z}}P  +{\it R_2}  Q \right) \,,
\end{equation}
\begin{equation} \label{eq25}
\tilde {Y}  = 
{\frac {\partial }{\partial z}}Y
+ \left( {\it R_1}  - h_{{z}}  \right) Y 
+{{\rm e}^{h }}{\it R_2}Q  \,.
\end{equation}
Taking into account the relation \eqref{eq2}, the equations \eqref{eq5}, \eqref{eq6} for $Q$
take the form
\begin{equation} \label{eq26}
{{\rm e}^{-h }}
\left( {\it Y_{r}}+ { \it h_{r}}\,Y+{\frac {1}{r}}\,Y \right)   -{\it Q_{z}} =0 \,,
\end{equation}
\begin{equation} \label{eq27}
{{\rm e}^{-h }}
\left( {\it Y_{z}}+ { \it h_{z}}\,Y \right)   +{\it Q_{r}} =0 \,.
\end{equation}

\section{Application of the nonlocal Darboux transformation }

The equations \eqref{eq22}, \eqref{eq23} contain unknown function $s$ 
and function $h$ associated with the initial potencial $u$ by the formula  \eqref{eq4}.
According to the formula \eqref{eq11}, we can take $h_u=-\ln  \left( rf \left( r,z \right)  \right)$
where $f$ is any solution of the Schr\"odinger equation  \eqref{eq1} with the initial potential $u$.

Let us consider 
$u=0, \, {\it f}={\frac {1}{\sqrt {{r}^{2}+{z}^{2}}}}$ and 
$h_0=-\ln  \left( r \right) +{\frac{1}{2}}\,\ln  \left( {r}^{2}+{z}^{2} \right) $ respectively.
An example of the particular solution of equations \eqref{eq22}, \eqref{eq23} can be
found for the ansatz $s=S \left( {r}^{2}+{z}^{2} \right)$. The equation \eqref{eq22} is satisfied
for any function $S$. The equation \eqref{eq23} leads to the solution
\begin{multline} \label{eq28}
s_0  = -1/2\,\ln  \left( {r}^{2}+{z}^{2} \right)
+\ln  \left(  \left( {r}^{2}+{z}^{2} \right) ^{{\it C_1}}+C \right)
\\
 -\ln  \left(  \left( 1-2\,{\it C_1} \right)  \left( {r}^{2}+{z}^{2} \right) ^{{\it C_1}}
+ \left( 1+2\,{\it C_1} \right) C \right)  \,,
\end{multline}
where $C, C_1$ are arbitrary constants. 
By the formula \eqref{eq4} 
with $\tilde h = h_0+s_0$ we 
obtain the new potential
\begin{equation} \label{eq29}
\tilde {u} \left( r ,z \right) = 
-8\,{\frac {C{{\it C_1}}^{2} \left( {r}^{2}+{z}^{2} \right) ^{{\it C_1}-
1}}{ \left(  \left( {r}^{2}+{z}^{2} \right) ^{{\it C_1}}+C \right) ^{2}}} \,.
\end{equation}
This potential satisfies the condition $\tilde {u}<0$ and has no singularities,
provided that $C>0, C_1 \geq 1$. Thus, we obtained two-parameter family
of solvable Helmholtz potentials.

From the formulas \eqref{eq25}-\eqref{eq27} we have the formula
for the solution of the equation  \eqref{eq1} with the potential \eqref{eq29} 
\begin{multline} \label{eq30}
\tilde {Y} = 
{\frac {\partial }{\partial z}}Y
\\
-{\frac {
 \left(  \left( 1+2\,{\it C_1} \right)  \left( {r}^{2}+{z}^{2} \right) 
^{{\it C_1}}+C \left( 1-2\,{\it C_1} \right)  \right)  \left( zY -\sqrt {{r}^{2}+{z}^{2}}Q \right) }
{ 2\,\left( {r}^{2}+{z}^{2} \right)  \left(  \left( {r}^{2}+{z}^{2}
 \right) ^{{\it C_1}}+C \right) }} \,,
\end{multline}
where the function $Q$
is a solution of the system of equations 
\begin{equation} \label{eq31}
\left( {\it Y_{r}} +{\frac {r\,Y}{{r}^{2}+{z}^{2}}} \right) {\frac {r}{\sqrt {{r}^{2}+{z}^{2}}}}
-{\it Q_{z}} =0 \,,
\end{equation}
\begin{equation} \label{eq32}
\left({\it Y_{z}} +{\frac {zY }{{r}^{2}+{z}^{2}}} \right){\frac {r}{\sqrt {{r}^{2}+{z}^{2}}}}
+{\it Q_{r}} =0 \,,
\end{equation}
and ${ \it Y}$ is any solution of equation \eqref{eq1} with the initial potential $u=0$.

For example let us consider the following simple solutions of equation \eqref{eq1}:
$1,z,{r}^{2}-2\,{z}^{2},3\,z{r}^{2}-2\,{z}^{3},{\frac {1}{\sqrt {{r}
^{2}+{z}^{2}}}},\ln  \left( r \right)$. Substituting this solutions into the equations
\eqref{eq30}-\eqref{eq32} we obtain the corresponding solutions of equation \eqref{eq1} with
the potential \eqref{eq29}:
\begin{equation*}
{\tilde {Y}}_{1} =
{\frac { \left( 1+2\,{\it C_1} \right)  \left( {r}^{2}+{z}^{2} \right) 
^{{\it C_1}}+ \left( 1-2\,{\it C_1} \right) C}{ \sqrt {{r}^{2}+{z}^{2}} \left(  \left( {r}^{2}+{
z}^{2} \right) ^{{\it C_1}}+C \right)}} \,,
\end{equation*}
\begin{equation*}
{\tilde {Y}}_{2} =
{\frac { \left( 1-2\,{\it C_1} \right)  \left( {r}^{2}+{z}^{2} \right) 
^{{\it C_1}}+ \left( 1+2\,{\it C_1} \right) C}{ \left( {r}^{2}+{z}^{2}
 \right) ^{{\it C_1}}+C}} \,,
\end{equation*}
\begin{equation*}
{\tilde {Y}}_{3} =
{\frac {z \left(  \left( 3-2\,{\it C_1} \right)  \left( {r}^{2}+{z}^{2}
 \right) ^{{\it C_1}}+ \left( 3+2\,{\it C_1} \right) C \right) }{
 \left( {r}^{2}+{z}^{2} \right) ^{{\it C_1}}+C}} \,,
\end{equation*}
\begin{equation*}
{\tilde {Y}}_{4} =
{\frac { \left({r}^{2} -2\,{z}^{2} \right)  \left(  \left(5 -2\,{\it 
C_1} \right)  \left( {r}^{2}+{z}^{2} \right) ^{{\it C_1}}+ \left(5+2\,{
\it C_1} \right) C \right) }{ \left( {r}^{2}+{z}^{2} \right) ^{{\it 
C_1}}+C}} \,,
\end{equation*}
\begin{equation*}
{\tilde {Y}}_{5} =
{\frac {z \left(  \left( 3+2\,{\it C_1} \right)  \left( {r}^{2}+{z}^{2}
 \right) ^{{\it C_1}}+ \left( 3-2\,{\it C_1} \right) C \right) }{
 \left( {r}^{2}+{z}^{2} \right) ^{3/2} \left(  \left( {r}^{2}+{z}^{2}
 \right) ^{{\it C_1}}+C \right) }} \,,
\end{equation*}
\begin{equation*}
{\tilde {Y}}_{6} =
{\frac { \left(  \left( 1+2\,{\it C_1} \right)  \left( {r}^{2}+{z}^{2}
 \right) ^{{\it C_1}}+ \left( 1-2\,{\it C_1} \right) C \right)  \left( 
\ln  \left( z+\sqrt {{r}^{2}+{z}^{2}} \right) -\ln  \left( r \right) 
 \right) }{ \left(  \left( {r}^{2}+{z}^{2} \right) ^{{\it C_1}}+C
 \right) \sqrt {{r}^{2}+{z}^{2}}}} \,.
\end{equation*}

Now the potential \eqref{eq29} can be taken as the initial potential for new 
transformations. In the case of the two - dimensional Schr\"odinger equation 
it was shown that the twofold application of the Moutard 
transformation can be effective for obtaining nonsingular 
potentials in cartesian coordinates  \cite {Tsarev2008} and
in cylindrical coordinates  \cite {Kudryavtsev2020}.
To avoid cumbersome formulas let us consider the simple
case $C_1=1$. The initial potential has the form
\begin{equation} \label{eq33}
u=-{\frac {8\,C}{ \left( {r}^{2}+{z}^{2}+C \right) ^{2}}} \,.
\end{equation}
For the first example let us take ${\tilde {Y}}_{1}$ and ${\tilde {Y}}_{2}$ 
at  $C_1=1$ as the solutions of the initial equation:
\begin{equation*}
{\it Y_1}=
{\frac {3\,\left({r}^{2}+{z}^{2}\right)-C}
{ \sqrt {{r}^{2}+{z}^{2}}\left( {r}^{2}+{z}^{2}+C \right)  
}}
\,\,, \,
{\it Y_2}=
{\frac {{r}^{2}+{z}^{2}-3\,C}{{r}^{2}+{z}^{2}+C}} \,.
\end{equation*}
From the equations \eqref{eq19}-\eqref{eq20} we obtain 
\begin{equation*}
{\it F}=
{\frac {z}{\sqrt {{r}^{2}+{z}^{2}}}}+K \,,
\end{equation*}
where $K$ is an arbitrary constant. Then from the formula  \eqref{eq18}
we obtain new solvable potential
\begin{equation} \label{eq34}
{\tilde {\tilde {u}}}  =
-{\frac {8\,C}{ \left( {r}^{2}+{z}^{2}+C \right) ^{2}}}+{\frac { 2\,\left(Kz+
\sqrt {{r}^{2}+{z}^{2}} \right) }{ \left( z+\sqrt {{r}^{2}+{z}^{2}}K \right) ^{
2}\sqrt {{r}^{2}+{z}^{2}}}} \,\,.
\end{equation}

For the second example let us take ${\tilde {Y}}_{2}$ and ${\tilde {Y}}_{3}$ 
at  $C_1=1$ as the solutions of the initial equation:
\begin{equation*}
{\it Y_1}=
{\frac {{r}^{2}+{z}^{2}-3\,C}{{r}^{2}+{z}^{2}+C}}
\,\,, \,
{\it Y_2}=
{\frac {z \left( {r}^{2}+{z}^{2}+5\,C \right) }{{r}^{2}+{z}^{2}+C}}  \,.
\end{equation*}
From the equations \eqref{eq19}-\eqref{eq20} we obtain 
\begin{equation*}
{\it F}=
{\frac {{r}^{4}+ \left({z}^{2} -15\,C \right) {r}^{2}}{{r}^{2}+{z}^
{2}+C}}
+K  \,,
\end{equation*}
where $K$ is an arbitrary constant. Then from the formula  \eqref{eq18}
we obtain new solvable potential
\begin{equation} \label{eq35}
{\tilde {\tilde {u}}}  =
4\,  {\it N}/ \left( {r}^{4}+ \left( {z}^{2}+K-15\,C
 \right) {r}^{2}+K \left( {z}^{2}+C \right)  \right) ^{2}  \,,
\end{equation}
where
\begin{multline*}
{\it N}=
 \left({r}^{2} -K \right)  \left( {r}^{2}+{z}^{2} \right) 
^{2}-C \left( 30\,{z}^{2}+22\,K-225\,C \right) {r}^{2}
\\
+KC \left( 14\,{
z}^{2}-2\,K+15\,C \right)  \,. 
\end{multline*}
This potential has no singularities, provided that $C>0, K \geq 15\,C$. 

\section{Results and Discussion}

The stationary Schr\"odinger and Helmholtz equations 
in the case of axial symmetry are investigated.
Using the approach of the papers 
\cite {Kudryavtsev2013}, \cite {Kudryavtsev2016}
the nonlocal Darboux transformations of the two - dimensional
stationary Schr\"odinger and Helmholtz equations in cylindrical coordinates  
are considered. Formulae for the nonlocal Darboux transformation 
are obtained and its relation to the generalized Moutard transformation is 
established.  New examples of two - dimensional  
potencials and exact solutions for the stationary axially symmetric 
Schr\"odinger and Helmholtz equations  are obtained as an application of 
the nonlocal Darboux transformation.
The examples considered show that by combining the
nonlocal Darboux transformation and the generalized Moutard transformation,
one can obtain many new examples of solvable potentials and exact
solutions for stationary axially symmetric Schr\"odinger and Helmholtz
equations.


\begin{thebibliography}{99}

\bibitem{Landau}
{L.D. Landau and E.M. Lifshitz, \em Quantum Mechanics:
Nonrelativistic Theory,}
 Pergamon Press, 1977.

\bibitem{Morse}
{P.M. Morse and K.U. Ingard, \em Theoretical Acoustics,} New York, NY:
McGraw- Hill, 1968.

\bibitem{Vinogradova}
{M. B. Vinogradova, O. V. Rudenko and A. P. Sukhorukov, \em 
The Wave Theory,} Nauka Publishers, Moscow, 1990.

\bibitem{Kak}
A.C. Kak and M. Slaney, Principles of Computerized Tomographic
Imaging, Society of Industrial and Applied Mathematics, 2001.

\bibitem{Novikov2010}
P.G. Grinevich, A.E. Mironov and S.P. Novikov, 2D-Schrodinger
Operator, (2+1) evolution systems and new reductions, 2D-Burgers
hierarchy and inverse problem data, arXiv:1005.0612 and Russian
Math Surveys, 2010, v 65, n 3, p. 580-582.

\bibitem{Matveev1991}
{V.B. Matveev, M.A. Salle, \em Darboux Transformations and
Solitons,} Springer, 1991.

\bibitem{Athorne 1991}
C. Athorne and J. J. C. Nimmo, On the Moutard transformation for 
integrable partial differential equations, Inverse Problems, 7 (1991), p.
809–826.

\bibitem{Andrianov 2012}
A A Andrianov and M V Ioffe, Nonlinear Supersymmetric Quantum 
Mechanics: concepts and realizations, 2012 J. Phys. A: Math. Theor. 45
(2012), p. 503001

\bibitem{Ioffe  2020}
M. V. Ioffe and D. N. Nishnianidze, 
Generalization of SUSY intertwining relations: New exact solutions of 
Fokker-Planck equation, EPL 129 (2020),p. 61001

\bibitem{Kudryavtsev2013}
A.G. Kudryavtsev, Exactly solvable two - dimensional stationary
Schr\"odinger operators obtained by the nonlocal Darboux
transformation, Phys. Lett. A, 377 (2013) 2477-2480.

\bibitem{Kudryavtsev2016}
A.G. Kudryavtsev, Nonlocal Darboux transformation of the
two-dimensional stationary Schr\"odinger equation
and its relation to the Moutard transformation, Theoretical 
and Mathematical Physics, 187(1)  (2016) 455-462.

\bibitem{Ibragimov1991}
I.S. Akhatov, R.K. Gazizov and N.H. Ibragimov. Nonlocal
symmetries. Heuristic approach. J. Sov. Math. 55 (1991) 1401-1450.

\bibitem{Bluman2010}
{G.W. Bluman, A.F. Cheviakov and S.C. Anco, \em Applications of
Symmetry Methods to Partial Differential Equations,}
 Springer, 2010.

\bibitem{Vinogradov 1999}
I.S. Krasil’shchik, and A.M. Vinogradov (Eds.) “Symmetries and conservation laws for differential
equations of mathematical physics” (1999). Translations of Mathematical Monographs
Vol. 182, AMS, Providence.

\bibitem{Kudryavtsev2020}
A.G. Kudryavtsev, Exact solutions of the time-independent 
axially symmetric Schr\"odinger equation, JETP Letters, 2020, Vol. 111, 
No. 2 (2020), pp. 126–128.

\bibitem{Risken1989}
{H. Risken, \em The Fokker-Planck Equation,}
 Springer, 1989.

\bibitem{Tsarev2008}
I.A. Taimanov, S.P. Tsarev, Two-dimensional rational solitons
and their blowup via the moutard transformation, Theoretical and
Mathematical Physics, November 2008, Volume 157, Issue 2, pp
1525-1541.


\end{thebibliography}
\end{document}